\documentclass[12pt]{article}

\begin{document}

\sf
\centerline{\Huge Proton 
decay, supersymmetry}

\centerline{\Huge  breaking and its mediation}

\vspace{7mm}

\centerline{\large
Borut Bajc$^{1}$ and 
Goran Senjanovi\'c$^{2}$}
\vspace{1mm}
\centerline{$^{1}$ {\it\small J.\ Stefan Institute, 1001 Ljubljana, Slovenia}}
\centerline{$^{2}${\it\small International Centre for Theoretical Physics,
Trieste, Italy}}

\vspace{5mm}
   
\centerline{\large\sc Abstract}
\begin{quote}
\small

We study the breaking of supersymmetry and its transmission to the light 
states in the context of the minimal SU(5) grand unified theory, with no 
additional singlets. This simple theory can be taken as a prototype for a 
program of breaking simultaneously grand unified symmetry and 
supersymmetry. The main predictions are: (i) d=6 proton decay is 
completely negligible and d=5 is in accord with experiment, (ii) 
supersymmetry breaking is mainly mediated by gravity.


\end{quote}
\rm

\section{Introduction}

After more than 30 years of supersymmetry playing a prominent role in 
particle physics we still know nothing about the source of its breaking or 
the nature of its mediation to the standard model supermultiplets. The 
most appealing scenario of spontaneous supersymmetry breaking in the MSSM 
fails by predicting sfermions lighter than the light fermions 
\cite{Dimopoulos:1981zb} and so the desired spontaneous breaking is 
assumed to happen in the SM gauge invariant sector and then transmitted to 
our world through either gravity or other interactions.

The most natural messengers of supersymmetry breaking are the Higgs 
doublets, $H$ and $\overline{H}$, as suggested some 10 years ago by 
Dvali 
and Shifman \cite{Dvali:1996hs}. Unfortunately this gives a negative 
contribution (proportional to the square of the Yukawa couplings $y^\dagger 
y$) to the squares of sfermion masses, so that the stop becomes tachyonic 
\cite{Dine:1996xk}.

In a sense this is a blow to the whole program. After all, the large $y_t$ 
plays an important role in supersymmetry for it leads naturally to the 
tachyonic property of the Higgs \cite{Inoue:1982pi} and it was also 
predicted \cite{Marciano:1981un} originally in order to achieve 
unification of couplings in the MSSM 
\cite{Dimopoulos:1981yj,Marciano:1981un}. New vectorlike multiplets can be 
added in order to mediate supersymmetry breaking but this typically means 
introducing new Yukawa type couplings \cite{Giudice:1998bp}. One pretends 
that they are zero and speaks euphemistically of pure gauge mediation, but 
this is true only if the gauge quantum numbers don't allow direct Yukawas, 
which is rare. Recently it was argued that the job could be done by the 
Higgs \cite{Joaquim:2006uz} or gauge \cite{Dermisek:2006qj} fields of some 
grand unified theory. In view of nonvanishing neutrino masses a particular 
interesting candidate is the Higgs supermultiplet responsible for the type 
II seesaw \cite{Joaquim:2006uz}. The crucial issue here is to know who 
dominates the mediation and by how much. This can be only answered in a 
simple and predictive theory, a kind we describe below.

Whoever the messenger is, an important question remains regarding the 
source of supersymmetry breaking. The conventional perturbative scenarios 
which use gauge singlets work kind of trivially due to the absence of 
constraints on the singlet couplings. Low energy supersymmetry has its 
principal role in grand unified theories, where it protects the Higgs from 
the large scale once the doublet-triplet (DT) splitting is achieved. Thus 
the most natural source of supersymmetry breaking is provided by the GUT 
Higgs supermultiplet, the SM singlet component. It turns out that this was 
studied very little \cite{Ovrut:1984qj,Drees:1985jx,Agashe:1998kg}.

At the same time supersymmetric grandunification is normally plagued by 
large threshold effects which impede precise predictions of the proton 
decay rate. For example, in the minimal supersymmetric SU(5) the dominant 
d=5 operator depends crucially on the ratio of the colour octet 
($\sigma_8$) mass $m_8$ and the weak triplet ($\sigma_3$) mass $m_3$ of 
the surviving remnants of the adjoint Higgs: varying $m_3/m_8$ from 1 to 4 
increases $\tau_p$ by a factor of $10^3$ \cite{Bachas:1995yt,Bajc:2002bv}. 
Furthermore, in general even soft supersymmetry breaking may obscure 
proton decay predictions, if the soft terms in the heavy and light sector 
are strongly split (for recent work see \cite{Berezhiani:2006mt} and 
references therein).

All of this indicates that by itself none of the above questions can be 
easily answered. It is strongly suggestive that our best hope is a 
consistent correlated treatment of all the three questions above 
(mediation, supersymmetry breaking and unification) in the context of a 
well defined simple grand unified theory. This is the main scope of our 
paper. For the sake of simplicity, clarity and predictivity, we discuss 
this program in the very minimal supersymmetric SU(5) theory. By this we 
mean besides the usual generations of quarks and leptons only $24_H$ and 
$5_H$, $\overline{5}_H$ supermultiplets. Of course the already existing 
phenomenology requires the inclusion of higher dimensional terms. 

An additional issue to be faced in SU(5) is the neutrino mass. Here there 
are number of ways which basically do not change anything we do in this 
paper. One simple possibility is for example bilinear R-parity breaking 
\cite{Kaplan:1999ds} 
(this means tuning away the baryon number violating contribution) 
which does not require any change. Another simple possibility is to have 
righthanded neutrinos as SU(5) singlets and utilize the so-called type I 
seesaw mechanism \cite{seesaw}. As long as these fields have zero vacuum 
expectation 
value and zero F-term, everything we say here goes through unchanged. In 
the opposite case one faces a danger of having potentially uncontrollable 
R-parity breaking which we prefer to avoid. Yet another simple possibility 
is to utilize type II seesaw \cite{Magg:1980ut} 
through the introduction of $15_H$ and 
$\overline{15}_H$ fields. These fields are potential messengers of 
supersymmetry breaking and we will comment on their role in section 
\ref{messenger}. Finally, one can use the triplet and singlet fermions 
in $24_H$ as a combination of type I and type III seesaw \cite{Ma:1998dn}.

We start by readdressing the issue of supersymmetry breaking through a 
single $24_H$ field in the supergravity potential. We find that this 
program can 
lead to a huge suppression of dimension 5 proton decay rate, due to the 
automatic appearence of intermediate states. In the case of the simplest 
possible realistic superpotential (quartic in $24_H$), this is actually a 
firm prediction. 

The possible mediators of supersymmetry breaking are:
1) gravity; 
2) heavy gauge bosons $X$ and $Y$; 
3) heavy Higgs supermultiplets $\sigma_3$ and $\sigma_8$ (weak triplet and 
colour octet from $24_H$), 4) $T$, $\overline{T}$ (the colour triplets 
from $5_H$ and $\overline{5}_H$ which mediate proton decay); 5) light 
Higgs doublets $D$ and $\overline{D}$. Since the masses of these states 
are constrained by the requirement of unification, one can compare their 
contribution to the soft light spartner masses. This program is rather 
predictive: as we show below, the dominant contribution to the soft 
breaking terms comes actually from gravity in most of the parameter space. 
The desired gauge mediation is rather suppressed and the question of 
flavour violation of neutral currents remains still an open question. 
However, in a rather small region of the parameter space, where $m_{3,8}$ 
are particularly fine-tuned, $\sigma_{3,8}$ could be the dominant 
messengers. The interesting characteristic of this case would be a 
somewhat unusual spectrum of spartners with right-handed sleptons much 
lighter than the rest.

Needless to say, we do not wish to argue here that this is the final 
theory, but rather to indicate how a well defined approach of using a 
simple model makes simultaneously clear predictions on proton decay, the 
TeV effective theory and the nature of the soft supersymmetry breaking.

In summary, the main predictions for the reader to carry away from this 
approach are:

\noindent
a) the minimal nonrenormalizable SU(5) with $24_H$, $5_H$, 
$\bar 5_H$ and three generations of $10_F$, $\bar 5_F$ 
embedded in supergravity is enough to break supersymmetry 
and get a realistic low-energy physics (MSSM); 

\noindent 
b) $d=6$ proton decay is completely negligible and $d=5$ is not 
in contradiction with the experiment as often claimed; the importance of 
this result cannot be overstressed;

\noindent
c) gravity dominates soft supersymmetry breaking terms in most of the 
parameter space, and if it were to be subdominant, the righthanded 
sleptons become the lightest spartners. 

\section{Breaking supersymmetry by ${\bf 24_H}$} 
\label{sec1}

As an example that is enough generic and illustrative, but still simple, 
we will consider the superpotential up to the fifth order \footnote{We 
will comment on the more restrictive cubic and quartic superpotential at 
the end of this section.} in the adjoint $24_H$ ($\Sigma$) and up to an 
arbitrary constant $W_0$

\begin{eqnarray}
\label{superpotential}
W-W_0=
&+&a_0\frac{v^3}{M_{Pl}^2}Tr\Sigma^2
+a_1\frac{v^2}{M_{Pl}^2}Tr\Sigma^3\nonumber\\
&+&a_2^{(1)}\frac{v}{M_{Pl}^2}Tr\Sigma^4
+a_2^{(2)}\frac{v}{M_{Pl}^2}\left(Tr\Sigma^2\right)^2\nonumber\\
&+&a_3^{(1)}\frac{1}{M_{Pl}^2}Tr\Sigma^5
+a_3^{(2)}\frac{1}{M_{Pl}^2}Tr\Sigma^3Tr\Sigma^2\;,
\end{eqnarray}

\noindent
where $v$ ($=M_{GUT}$) is the grand unified scale and $M_{Pl}$ stands 
for the Planck scale ($\approx 10^{19}$ GeV). The reader should keep 
in mind that some of the coefficients $a_i$ (except the last two) 
could be bigger than $1$ without being in contradiction with 
perturbativity. 

One expands the $\Sigma$ multiplet as

\begin{eqnarray}
\Sigma=&+&\frac{\sigma}{\sqrt{30}}\;{\rm diag}(2,2,2,-3,-3)\nonumber\\
&+&\frac{\sigma_8}{\sqrt{2}}\;{\rm diag}(1,-1,0,0,0)+
\frac{\sigma_3}{\sqrt{2}}\;{\rm diag}(0,0,0,1,-1)
\end{eqnarray}

\noindent
with $\sigma$'s canonically normalized, so that the K\" ahler 
potential is just 

\begin{equation}
\label{kahler}
K=Tr\Sigma^\dagger\Sigma\;.
\end{equation}

We take here the canonical K\" ahler only for simplicity, although it is 
not realistic. A more general case would only help to achieve 
supersymmetry breaking. Of course, we will not take seriously any 
predicition that the minimal K\" ahler leads to, such as for example 
flavour conservation in neutral currents at high energies. Nothing in the 
discussion below depends on this assumption.

It is easy to check that in supergravity $\langle\sigma_{3,8}\rangle =0$ 
is an extremum. We will see soon that the supersymmetric mass of the weak 
triplet and color octet is larger than the supersymmetry breaking ones, so 
the solution is at least locally stable (up to possible tunneling). By 
definition, $\langle\sigma\rangle =v$ and we look for nonvanishing $F$ in

\begin{equation}
\sigma=v+\theta\theta F\;.
\end{equation}

Now everything reduces to the minimization of the supergravity 
potential with the superpotential and (canonical) Kahler potential

\begin{eqnarray}
\label{w}
W-W_0&=&\sum_{n=0}^{3}b_n\frac{v^{3-n}}{M_{Pl}^2}\sigma^{n+2}\;,\\
\label{k}
K&=&\sigma^*\sigma\;.
\end{eqnarray}

The coefficients $b$'s are expressed as

\begin{eqnarray}
b_0&=&a_0\;,\\
b_1&=&-\frac{1}{\sqrt{30}}a_1\;,\\
b_2&=&\frac{7}{30}a_2^{(1)}+a_2^{(2)}\;,\\
b_3&=&-\frac{13}{30\sqrt{30}}a_3^{(1)}-\frac{1}{\sqrt{30}}a_3^{(2)}\;.
\end{eqnarray}

From (\ref{w}) it is easy to calculate the various derivatives; 
the system obtained is linear in the couplings $b$'s:

\begin{eqnarray}
\pmatrix{
  1
& 1
& 1
& 1
\cr
  2
& 3
& 4
& 5
\cr
  2
& 6
& 12
& 20
\cr
  0
& 6
& 24
& 60
\cr}
\pmatrix{
b_0
\cr
b_1
\cr
b_2
\cr
b_3
\cr}=
M_{Pl}^2\pmatrix{
\left(W-W_0\right)/v^5
\cr
W'/v^4
\cr
W''/v^3
\cr
W'''/v^2
\cr}\;.
\end{eqnarray}

This system is easily inverted to get 

\begin{eqnarray}
\pmatrix{
b_0 
\cr
b_1
\cr
b_2
\cr
b_3
\cr}=
M_{Pl}^2\pmatrix{
  10
& -6
& 3/2
& -1/6
\cr   
  -20
& 14
& -4
& 1/2   
\cr
  15
& -11
& 7/2
& -1/2
\cr
  -4
& 3
& -1
& 1/6
\cr}
\pmatrix{
\left(W-W_0\right)/v^5
\cr
W'/v^4
\cr
W''/v^3
\cr   
W'''/v^2
\cr}\;.
\end{eqnarray}

The last step is to find out for which $W$, $W'$, $W''$ and $W'''$ 
is the vev $v$ a stable minimum of the supergravity potential

\begin{equation}
V=\exp{\left(K/M_*^2\right)}\left[
\left(
\frac{\partial W}{\partial \phi^i}+
\frac{\partial K}{\partial \phi^i}\frac{W}{M_*^2}\right)
\left(K^{-1}\right)^i_j
\left(
\frac{\partial W^*}{\partial \phi^*_j}+
\frac{\partial K}{\partial \phi^*_j}\frac{W^*}{M_*^2}\right)
-3\frac{\left| W\right|^2}{M_*^2}\right]\;,
\end{equation}

\noindent
where $M_*=M_{Pl}/\sqrt{8\pi}\approx 2\times 10^{18}$ GeV 
is the so called reduced Planck mass and $\left(K^{-1}\right)^i_j$ 
is the inverse matrix of $\partial^2 K/\partial\phi^i\partial\phi^*_j$.
The fine-tuning of the cosmological constant requires $V=0$ at the 
minimum $\sigma=v$. Together with the constraint of the minimum 
($dV/d\sigma=0$ and all scalar masses square positive) this 
after some calculation leads to

\begin{eqnarray}
\label{wp}
\frac{W'}{W}&=&\frac{\sqrt{3}\eta^*}{M_*}-\frac{v^*}{M_*^2}\;,\\
\label{wpp}
\frac{W''}{W}&=&\left(\frac{\sqrt{3}\eta^*}{M_*}-\frac{v^*}{M_*^2}\right)^2-
\left(\frac{\eta^*}{M_*}\right)^2\;,\\
\label{wppp}
\frac{W'''}{W}&=&3\left(\frac{\sqrt{3}\eta^*}{M_*}-\frac{v^*}{M_*^2}\right)
\left(\left(\frac{\sqrt{3}\eta^*}{M_*}-\frac{v^*}{M_*^2}\right)^2-
\left(\frac{\eta^*}{M_*}\right)^2\right)\nonumber\\
&-&2\left(\frac{\sqrt{3}\eta^*}{M_*}-\frac{v^*}{M_*^2}\right)^3
+\frac{2\eta^{*3}}{\sqrt{3}M_*^3}\left(1+\xi\right)\;,
\end{eqnarray}

\noindent
where $\left|\eta\right|^2=1$ and $\left|\xi\right|\le 1$ 
($|\xi|=1$ means one massless scalar).

The above equations tell us that effectively the first three derivatives 
of the superpotential must be highly fine tuned, i.e. the field $\sigma$ is 
very close to a flat direction, as physically expected if gravity is to 
play a substantial role.

Finally, by definition the gravitino mass is 

\begin{equation}
\label{m32}
m_{3/2}=\frac{|W|}{M_*^2}e^{K/M_*^2}\;.
\end{equation}

In principle $m_{3/2}$ can be fine-tuned to be as small as one wants, 
but in low energy supersymmetry it is expected to lie around TeV. 

The parameter $W_0$ is only constrained to satisfy the upper bound 
$v^5/M_{Pl}^2$, which comes from the requirement $b_3\le 1$. Its value is 
locally (close to our minimum) completely irrelevant, but plays an 
important role for the global shape of the potential. For example, besides 
our local minimum with vanishing energy there is at least one more minimum 
to worry about, i.e. $\langle\sigma\rangle=0$, whose energy is given by

\begin{equation}
E\left(\langle\sigma\rangle=0\right)=-3\frac{|W_0|^2}{M_*^2}\;.
\end{equation}

Of course there could be other local minima, depending on the value of 
$W_0$. For example, for $W_0=0$ the closest minimum to ours (and lower in 
energy) lies at approximately $1.4 M_{GUT}$. One is thus faced with an 
important question of metastability of our local minimum. The tunneling to 
the ground state turns out to be very slow \cite{Duncan:1992ai} as 
expected from the large distance between the minima.

The above discussion is both simple and generic enough to illustrate all 
the essential points of this program. Still, one may ask, why not a 
simpler superpotential. The cubic case can be disposed of immediately, 
since it has only two couplings ($a_0$ and $a_1$ in 
(\ref{superpotential})), insufficient to satisfy (\ref{wp})-(\ref{wppp}).

The quartic case on the other hand can suffice, since it adds two new 
couplings ($a_2^{(1)}$ and $a_2^{(2)}$ as seen from 
(\ref{superpotential})). It leads to interesting predictions

\begin{equation}
W_0={\cal O}(m_{3/2}M_*^2)\;\;,\;\;
m_3=4m_8\;\;.
\end{equation}

These predictions are to be taken with a grain of salt, since they demand 
ignoring a number of higher dimensional operators. Still, it is 
interesting that the latter prediction automatically suppresses 
sufficiently the $d=5$ proton decay, as discussed in section \ref{rge}. 
The tunneling is still under control, and formally, although not strongly 
motivated, this case cannot be ruled out.

\section{The particle spectrum and DT splitting}

After the SU(5) symmetry breaking, the surviving elements of $24_H$, the 
SU(3) octet $\sigma_8$ and the SU(2) triplet $\sigma_3$ have the masses

\begin{equation}
\label{c3}
m_3=c_3\frac{M_{GUT}^3}{M_{Pl}^2}\;,\;
\label{c8}
m_8=c_8\frac{M_{GUT}^3}{M_{Pl}^2}\;, 
\end{equation}

\noindent
where 

\begin{equation}
\label{c3c8}
c_3\approx\frac{2}{3}a_2^{(1)}+\frac{1}{\sqrt{30}}a_3^{(1)}\;,\;
c_8\approx\frac{8}{3}a_2^{(1)}-\frac{28}{3\sqrt{30}}a_3^{(1)}
\end{equation}

\noindent
after using the symmetry breaking constraints from the previous section. 

The situation with $5_H$ and $\overline{5}_H$ supermultiplets require 
additional fine-tuning as everybody knows. From the additional terms in 
the superpotential

\begin{equation}
\label{five}
W_5=m_5 \overline{5}_H5_H+
\sqrt{30}\beta_1\overline{5}_H\Sigma 5_H+
30\beta_2\overline{5}_H\frac{\Sigma^2}{M_{Pl}} 5_H
\end{equation}

\noindent
one finds for the doublet ($D$) and triplet ($T$) mass terms 

\begin{eqnarray}
\label{mud}
\mu_D&=&m_5-3\beta_1v+9\beta_2\frac{v^2}{M_{Pl}}\;,\\
\label{mut}
\mu_T&=&m_5+2\beta_1v+4\beta_2\frac{v^2}{M_{Pl}}\;.
\end{eqnarray}

To get the light Higgs mass $\mu_D={\cal O}(m_W)$ one needs to 
fine-tune the combination of parameters on the righthandside in 
(\ref{mud}). This gives for the triplet mass 
$\mu_T=5\beta_1v-5\beta_2v^2/M_{Pl}$. Since $T$ and $\overline{T}$ 
mediate the $d=5$ proton decay, these masses must be as large 
as possible and thus $\beta_i$ cannot be small, i.e. at least 
$\beta_2\approx 0.1-1$. This has a dramatic impact on supersymmetry 
breaking, as seen immediately from the last two terms in (\ref{five}). 
Namely, this implies a contribution of order $(-3\beta_1+18\beta_2v/M_{Pl})F$ 
to the off-diagonal Higgs mass term which, without fine-tuning, requires 
$F \le$ TeV in order not to destabilize the Higgs masses. Such a small 
$F$ can work only if the Higgs doublets $D$ and $\overline{D}$ are the 
dominant mediators of supersymmetry breaking, but as discussed above, it 
implies a tachyonic stop. The escape from this impasse is to apply 
a further constraint on the model parameters:
$|-3\beta_1+18\beta_2v/M_{Pl}|\ll 1$, so that the dangerous off-diagonal 
contribution to the Higgs doublet mass is at most ${\cal O}(m_W^2)$. 
To summarize, although at the prize of two fine-tunings, the 
minimal model survives all the phenomenological constraints. 

For those who do not like so many fine-tunings, there is a 
different option for the doublet-triplet splitting and the hiding 
of the singlet $\sigma$ from the light Higgs doublets. 
This can be accomplished in two different ways. 
The simplest realization is to add a pair of $50_H$ and 
$\overline{50}_H$ multiplets, which contain colour triplets, but no weak 
doublets. Through the couplings \cite{Ovrut:1984qj}

\begin{equation}
W_{50}=\frac{1}{M_{Pl}}24_H^2\left(\overline{5}_H 50_H+
\overline{50}_H 5_H\right) +\left(M_{50}+\Sigma+...\right)
\overline{50}_H 50_H
\end{equation}

\noindent
(the dots stand for possible higher dimensional operators) one makes the 
triplets heavy and the doublets remain massless (until supersymmetryy gets 
broken). Clearly, the maximum mass the triplets can have is 
${\cal O}(M_{GUT}^2/M_{Pl})$. 

An alternative is to use $75_H$ \cite{Masiero:1982fe} instead of 
$24_H$, since $75_H$ behave as $24_H^2$ in the above example. It has 
direct renormalizable couplings and in this case $M_T\approx M_{GUT}$.

\section{RGE for gauge couplings: unification and proton decay}
\label{rge}

We start here with a careful discussion of the minimal supersymmetric 
SU(5) unification constraints independent of our program. A consistent 
renormalization group analysis assumes that the three masses $m_T$, $m_3$ 
and $m_8$ are free. At the renormalizable tree level, $m_3=m_8$, but minimal 
supersymmetric SU(5) makes no sense without higher dimensional terms, 
since it predicts wrongly fermions masses. Once the higher dimensional 
terms are allowed, as in our example, $m_3$ and $m_8$ become arbitrary.

At the one loop level, the RGE's for the gauge couplings are (we ignore 
here for simplicity higher dimensional terms which split the gauge 
couplings at the grand unified scale through $\langle\Sigma\rangle\ne 0$; 
see section \ref{messenger})

\begin{eqnarray}
\label{alfa1} 
2\pi\left(\alpha_1^{-1}(M_Z)-\alpha_U^{-1}\right)&=&
-{5\over 2}\ln{\Lambda_{SUSY}\over M_Z}
+{33\over 5}\ln{M_{GUT}\over M_Z}
+{2\over 5}\ln{M_{GUT}\over m_T}\;,\\
2\pi\left(\alpha_2^{-1}(M_Z)-\alpha_U^{-1}\right)&=&
-{25\over 6}\ln{\Lambda_{SUSY}\over M_Z}
+\ln{M_{GUT}\over M_Z}
+2\ln{M_{GUT}\over m_3}\;,\\
\label{alfa3}
2\pi\left(\alpha_3^{-1}(M_Z)-\alpha_U^{-1}\right)&=&
-4\ln{\Lambda_{SUSY}\over M_Z}
-3\ln{m_8\over M_Z}
+\ln{M_{GUT}\over m_T}\;.
\end{eqnarray}

\noindent
From (\ref{alfa1})-(\ref{alfa3}) we obtain

\begin{eqnarray}
2\pi\left(3\alpha_2^{-1}-2\alpha_3^{-1}-\alpha_1^{-1}\right)&=&
-2\ln{\Lambda_{SUSY}\over M_Z}   
+{12\over 5}\ln{m_T\over M_Z}
+6\ln{m_8\over m_3}\;,\\
2\pi\left(5\alpha_1^{-1}-3\alpha_2^{-1}-2\alpha_3^{-1}\right)&=&
8\ln{\Lambda_{SUSY}\over M_Z}
+36\ln{(\sqrt{m_3m_8}M_{GUT}^2)^{1/3}\over M_Z}\;.
\end{eqnarray}

We stick here to low energy supersymmetry, i.e. we take 
$\Lambda_{SUSY}\approx M_Z$, as required by one-loop unification. 

This gives

\begin{eqnarray}
\label{mt}
m_T&=&m_T^0\left({m_3\over m_8}\right)^{5/2}\;,\\
\label{mgut}
M_{GUT}&=&M_{GUT}^0\left({M_{GUT}^0\over\sqrt{m_3 m_8}}\right)^{1/2}\;.
\end{eqnarray}

In the above equations the superscript $^0$ denotes the values in the case 
$m_3=m_8=M_{GUT}$. Taking $\alpha_1^{-1}=59$, $\alpha_2^{-1}=29.57$ and 
$\alpha_3^{-1}=8.55$

\begin{equation}
\label{mgut0}
m_{GUT}^0\approx 10^{16}\; {\rm GeV}\;.
\end{equation}

If one ignores higher-dimensional terms, one predicts $m_3=m_8$ and thus 
$m_T=m_T^0$. It is known that $m_T^0$ is not large enough to bring $d=5$ 
proton decay in accord with experiment (unless one goes through painful 
gymnastics or arbitrary cancellations \cite{Bajc:2002bv}). At the same 
time, $M_{GUT}$ is obviously not predicted and can be as large as 
$10^{18}$ GeV, as long as $m_3\approx m_8\approx 10^{13}$ GeV (we stick to 
a perturbative theory and demand $M_{GUT}\le M_{Pl}/10$). This clearly 
requires a large amount of fine-tuning, since the mass of the SM singlet 
$\sigma$ must be about ten orders of magnitude smaller (recall that 
$m_\sigma ={\cal O}(m_{3/2}))$.

The intermediate values of $m_{3,8}$ on the other hand simply imply that 
the Yukawa $Tr\left(\Sigma^3\right)$ coupling is small. On the other hand, 
higher dimensional terms in the superpotential are the simplest 
possibility of curing wrong fermion mass relations in the theory; once 
they are included $m_3$ and $m_8$ become arbitrary, as in our case. This 
means that $m_T$ can be arbitrary large, and in what follows we demand 
$m_T\ge 10^{17}$ GeV in order to stabilize the proton.

The above message cannot be overstressed. We have argued that the theory 
does not predict either the GUT scale or the mass of the colour triplets, 
and we will need the experiment to learn their values. Instead of 
endlessly worrying about the nonexistent predictions of this prototype 
theory of supersymmetric grandunification, a correct procedure requires to 
take into account the whole parameter space without ad-hoc unphysical 
prejudices. The strong indication of large $m_T$ and thus $M_{GUT}$ 
requires only intermediate states $\sigma_3$ and $\sigma_8$, completely 
consistent with theory and experiment. Actually, simply demanding that 
supersymmetry be broken in the minimal scheme without any hidden sector 
implies automatically these intermediate states. The bottom line of all of 
this is that the dimension 6 proton decay operators can be completely 
ignored: $\tau_p(d=6)\approx 10^{40}$ yrs for $M_{GUT}\approx 10^{17}$ 
GeV.

In the minimal theory we considered, the triplet has a mass of 
order $M_{GUT}^2/M_{Pl}$. Due to the requirements of safe 
$d=5$ proton decay ($m_T\ge 10^{17}$ GeV) and 
$M_{GUT}\ll M_{Pl}$, the only possibility is to have 
$M_{GUT}\approx 10^{18}$ GeV. This determines the masses 
$m_3\approx 2 m_8\approx 10^{13}$ GeV as seen from (\ref{mt}) 
and (\ref{mgut}). 

We comment here on the alternatives that we mentioned in 
the previous section. 
If one wants to employ the missing partner mechanism, and 
give up the minimal model, there are more options. 
The simplest situation here is to consider $50_H$ and 
$\overline{50}_H$ as complete multiplets at $M_{GUT}^2/M_{Pl}$. 
It can be checked that this also guarantees no Landau pole 
below $M_{Pl}$. In the case of $75_H$, 
$m_T\approx M_{GUT}$. As there are more states which contribute 
to the increase of the gauge couplings, one is forced to have again 
$M_{GUT}\approx 10^{18}$ GeV in order to avoid a Landau pole below 
$M_{Pl}$.

The common characteristic of all the above cases is a large $M_{GUT}\approx 
10^{18}$ GeV, which completely suppresses $d=6$ proton decay, and makes it 
out of reach of even a future generation experiment. Dimension 5 proton 
decay is clearly in accord with experiment and in the last case above 
it may not be easily visible.

One may not be happy with such a high value of $M_{GUT}$, 
maybe too close to $M_{Pl}$. A possible way-out is to 
add another $24_H$ and stick to the fine-tuned doublet-triplet 
splitting. Clearly there are no other constraints here except 
for $M_{GUT}\ge m_T\ge 10^{17}$ GeV.

\section{Transmitting supersymmetry breaking}
\label{messenger}

As discussed in the introduction, there are a number of possible mediators 
of supersymmetry breaking: 1) gravity; 2) $X$ and $Y$ heavy vector 
supermultiplets; 3) heavy Higgs supermultiplets $\sigma_3$ and $\sigma_8$; 
4) heavy colour triplets $T$, $\overline{T}$ from $5_H$, $\overline{5}_H$ 
(and possibly $50_H$ and $\overline{50}_H$); 5) light Higgs doublets $D$ and 
$\overline{D}$.

All the supersymmetry breaking terms are necessarily proportional to $F$, 
which is the auxiliary field of the singlet supermultiplet 
$\sigma=v+\theta\theta F$. Due to the requirement of zero cosmological 
constant, it is connected to the gravitino mass

\begin{equation}
F\approx m_{3/2}M_*\;.
\end{equation}

We now carefully study each of these contributions. The end result will 
turn out to be the domination of gravity. For this reason we only present 
the estimates of the single contributions, i.e. the order of magnitude 
values for the soft terms.

\noindent
{\bf 1) Gravity}

Gravity is an automatic messenger in any theory, and its contribution to 
the sfermion masses and $A$-terms is

\begin{equation}
m_{\tilde {f}}\approx A\approx m_{3/2}\;. 
\end{equation}

The situation with gaugino masses depends on the following higher 
dimensional operator

\begin{equation}
\int d^2\theta \frac{f}{M_{Pl}} Tr\left(\Sigma W^\alpha W_\alpha\right)\;,
\end{equation}

\noindent
where $W^\alpha$ is the supersymmetric generalization of the Yang-Mills 
field strength. One gets generically for the gaugino masses 

\begin{equation}
m_\lambda\approx f m_{3/2}\;.
\end{equation}

If $f$ is of order $1$, the unification constraints must of course be 
reanalyzed. For smaller $f$ one expects lighter gauginos, a fact that 
helps further suppressing the $d=5$ proton decay. This encouraged us to 
focus on the case $f\ll 1$ in the above renormalization group study.

\noindent
{\bf 2) Heavy gauge bosons ${\bf X}$ and ${\bf Y}$}

In this case one gets for the soft terms the intuitively expected 
result \cite{Dermisek:2006qj}

\begin{equation}
m_{\tilde{f}}\approx m_\lambda\approx A\approx \frac{\alpha}{\pi} 
\frac{F}{M_{GUT}}\;.
\end{equation}

Since in this theory $M_{GUT}$ is expected to be near $M_*$, barring 
accidental cancellations involving complicated K\" ahler potentials, this 
contribution is negligible compared to gravity mediation.

\noindent
{\bf 3) Physical states in ${\bf 24_H}$: ${\bf \sigma_3}$ 
and ${\bf\sigma_8}$}

The contribution to the masses is given at two-loops, and is of the 
order (a similar contribution is also for the $A$ terms at one-loop)

\begin{equation}
m_{\tilde{f}}\approx A\approx \frac{\alpha}{\pi}\frac{F_i}{m_i}\;,
\end{equation}

\noindent
where $i=3$ and/or $8$ and 

\begin{equation}
F_i=F\left.\frac{\partial m_i}{\partial\sigma}\right|_{\sigma=v}\;.
\end{equation}

Typically $F_i/m_i={\cal O}(F/M_{GUT})$, which would make this 
contribution subdominant with respect to gravity, precisely because of the 
loop suppression. To overcome it one needs to fine-tune $m_i$ without 
suppressing at the same time $F_i$. In the model discussed here this 
reduces to fine-tune $c_3$ and $c_8$ in (\ref{c3})-(\ref{c3c8}) without the 
coefficients $a_2^{(1)}$, $a_3^{(1)}$ being much less than $1$. Since 
$m_3$ and $m_8$ must be of the same order of magnitude in order to prevent 
$m_T$ being much bigger than $M_{GUT}$ (see eq. (\ref{mt})), this is 
clearly impossible. Here $\sigma_3$ and $\sigma_8$ contribute no more than 
$X$ and $Y$.

The above is not a rigorous result, though. After all, one can include 
even higher dimensional terms in the superpotential in order to have the 
necessary freedom to fine-tune $m_3$ and $m_8$ to be small. Since at the 
same time one should keep $F_3$ and $F_8$ as large as possible, the ideal 
case is to stop at $\Sigma^6/M_{Pl}^3$. At first glance one could enhance 
arbitrarily the mediation of $\sigma_3$ and $\sigma_8$, but recall that

\begin{equation}
m_3\approx m_8 > 10^{12-13}\;{\rm GeV}
\end{equation}

\noindent
in order to keep $M_{GUT}$ below $M_{Pl}$. It is a simple exercise to 
check that, although this contribution can be made bigger than the 
one of $X$ and $Y$, it is at most of order $m_{3/2}$. In short, even 
after a fine-tuning, gravity still tends to dominate. 

\noindent
{\bf 4) Heavy colour triplets ${\bf T}$ and ${\bf \overline{T}}$}

Here the situation is very simple. Since these states must be rather 
heavy in order to stabilize the proton, their contribution, as in the case 
of $X$ and $Y$ is much smaller than the gravitational one. Similarly the 
possible contribution of $50_H$ and $\overline{50}_H$ states is also 
negligible since they lie at the GUT scale for the sake of unification and 
perturbativity up to $M_{Pl}$. 

\noindent
{\bf 5) Light Higgs}

As we discussed repeatedly, the light Higgs is never allowed to dominate, 
since it makes the stop tachyonic. Actually, in the cases when one splits 
the doublet and the triplet using the missing partner mechanism, light 
Higgses are completely decoupled from the source of supersymmetry 
breaking. In the opposite case, when one fine-tunes this coupling (the way 
one does for the $\mu$ term), the light Higgs contribution cannot be 
predicted, since it depends on the amount of fine-tuning. All one can say 
here is that the light Higgs cannot dominate.

A few words are needed regarding the issue of neutrino mass. As we said in 
the introduction, one possibility are the bilinear R-parity violating 
terms, which do not affect anything of the above. The same is true of the 
type I seesaw. The situation with the type II seesaw requires some 
discussion. The $15_H$ and $\overline{15}_H$ fields have been argued 
recently to be interesting messengers of supersymmetry breaking 
\cite{Joaquim:2006uz}. These fields are taken as complete multiplets at 
some intermediate scale in order not to affect the unification 
constraints. They couple to the adjoint and thus clearly transmit the 
supersymmetry breaking. In principle, with some fine-tuning they could be 
made to dominate the mediation of supersymmetry breaking. We prefer not to 
incorporate this case here seriously, for otherwise it requires an 
in-depth study of its impact on unification constraints and 
perturbativity. In any case the possibilty of these fields dominating 
supersymmetry mediation has been carefully studied in 
\cite{Joaquim:2006uz}.

As claimed, it is clear that in general no contribution except for gravity 
can be the dominant one. In any case, it is only $\sigma_3$ and $\sigma_8$ 
that can compete with gravity, which does not complicate things much, 
since gravity mediation makes no clear statements regarding the flavour 
structure of soft terms. It is worth emphasizing that the so called mSUGRA 
with universal soft terms does not emerge in supergravity since it is 
based on a completely unphysical assumption of canonical K\" ahler. In our 
study the assumption of a canonical K\" ahler was used only for simplicity 
and transparency and no prediction is based on it. We wanted to emphasize 
that generically supersymmetry can get broken by the $24_H$ once higher 
dimensional terms are allowed; non-canonical K\" ahler makes the task only 
easier.

In the extreme case of $\sigma_3$ and $\sigma_8$ being maximally 
fine-tuned and giving a somewhat bigger contribution than gravity, the 
singlet sleptons would be somewhat lighter ($\sigma_3$ and $\sigma_8$ 
carry no hypercharge). This is based on incomplete and rough estimates and 
it would have to be quantify in order to be taken very seriously. This 
task is beyond the scope of this letter, although it could be a useful 
exercise for the future.

\section{Conclusions}

The scenario of low energy supersymmetry is plagued by our complete 
ignorance of the source and the nature of supersymmetry breaking and its 
transmission to the spartners of the SM model particles. Perturbative 
approaches typically use gauge singlet fields to break supersymmetry which 
renders them prediction free. As we discussed in the introduction, there 
were important attempts, though, to use the GUT Higgs (the adjoint of 
SU(5)) to do the job, but with the price of introducing ad-hoc 
new fields. 

On the other hand, the mediation of the breaking, when not argued to be 
dominated by gravity, is typically attributed to new vectorlike states, 
introduced ad-hoc for this purpose. On top of that, one often ignores 
their possible Yukawa couplings and speaks of gauge mediation. Notable 
exceptions are the attempts to use the GUT gauge multiplets 
\cite{Dermisek:2006qj} and the SM model triplet responsible for the type II 
seesaw \cite{Joaquim:2006uz}.

In this paper we have studied supersymmetry breaking and its transmission 
to the light states in a simple grand unified theory such as $SU(5)$ 
without any ad-hoc singlets. The adjoint Higgs $24_H$ breaks the GUT 
symmetry and supersymmetry at the same time. While the SM 
gauge singlet direction must be quite flat, the color octet and the weak 
triplet end up at the intermediate scale; their impact on the running is 
to increase in general the GUT scale and possibly the masses of the color 
triplets states which mediate d=5 proton decay. We wish to emphasize again 
that this requires a large amount of fine-tuning, since the singlet 
$\sigma$ is at the TeV scale, while $\sigma_{3,8}$ are at intermediate 
scale of about $10^{13}$ GeV. The alternative would be adding more fields 
just to fix the unification constraints \cite{Ovrut:1984qj}. The bottom 
line: (1) the minimal theory $24_H$, $5_H$ and $\bar 5_H$ suffices; 
(2) d=6 proton decay gets out of reach; (3) d=5 is slowed enough to 
be in accord with the experimental limits.

The gauge structure of the theory (i.e. the absence of gauge singlets) 
makes it quite predictive even when it comes to the transmission of 
supersymmetry breaking to the MSSM particles. It turns out that gravity 
dominates in most of the parameter space, while, at the price of 
fine-tuning, the octet and the triplet of $24_H$ could compete with 
gravity. In the extreme and improbable situation of their domination, the 
signal would be the lightness of singlet sleptons. In short, this simple 
theory is an example of a predictive program of using grand unification to 
be responsible for breaking superymmetry and for the subsequent mediation 
without any new ad-hoc singlets whose existence makes the program both 
trivially achievable and prediction free. The generic prediction in this 
program is the existence of intermediate scale particles that push up the 
unification scale and keep the proton safe.

\section*{Acknowledgements}

We thank Gia Dvali and Alejandra Melfo for discussions and encouragement. 
The work of G.S. was supported in part by the European Commission under 
the RTN contract MRTN-CT-2004-503369; the work of B.B. was supported in 
part by the Slovenian Research Agency.

\end{document}